# Microscopic evidence for a chiral superconducting order parameter in the heavy fermion superconductor UTe$_2$

Lin Jiao[1], Sean Howard[1], Sheng Ran[2,3], Zhenyu Wang[1], Jorge Olivares Rodriguez[1], Manfred Sigrist[4], Ziqiang Wang[5], Nicholas P. Butch[2,3], Vidya Madhavan[1]

[1]Department of Physics and Frederick Seitz Materials Research Laboratory, University of Illinois Urbana-Champaign, Urbana, Illinois 61801, USA
[2]NIST Center for Neutron Research, National Institute of Standards and Technology, Gaithersburg, MD 20899, USA
[3]Center for Nanophysics and Advanced Materials, Department of Physics, University of Maryland, College Park, Maryland 20742, USA
[4] Institute for Theoretical Physics, ETH Zurich, 8093 Zurich, Switzerland
[5]Department of Physics, Boston College, Chestnut Hill, Massachusetts 02467, USA

**Spin-triplet superconductivity is a condensate of electron pairs with spin-1 and an odd-parity wavefunction[1]. A particularly interesting manifestation of triplet pairing is a chiral p-wave state which is topologically non-trivial and a natural platform for realizing Majorana edge modes[2,3]. Triplet pairing is however rare in solid state systems and so far, no unambiguous identification has been made in any bulk compound. Since pairing is most naturally mediated by ferromagnetic spin fluctuations, uranium based heavy fermion systems containing *f*-electron elements that can harbor both strong correlations and magnetism are considered ideal candidate spin-triplet superconductors[4-10]. In this work we present scanning tunneling microscopy (STM) studies of the newly discovered heavy fermion superconductor, UTe$_2$ with a $T_{SC}$ of 1.6 K[11]. We find signatures of coexisting Kondo effect and superconductivity which show competing spatial modulations within one unit-cell. STM spectroscopy at step edges show signatures of chiral in-gap states, predicted to exist at the boundaries of a topological superconductor. Combined with existing data indicating triplet pairing, the presence of chiral edge states suggests that UTe$_2$ is a strong candidate material for chiral-triplet topological superconductivity.**

Spin-triplet pairing is a fascinating example of an unconventional superconductivity, where the Cooper pairs have finite angular momentum and equal spin. More interestingly, in the special case of a chiral p-wave superconductor, theory predicts the existence of chiral Dirac quasiparticles at particular boundaries of the system. In principle, such a material could be the condensed matter testing ground for fundamental quantum field theories. The best-known example of chiral spin-triplet paring is the superfluid $^3$He-A phase[12] and over the last few decades, there has been an intensive search for potential spin-triplet superconductors in solid state systems. $Sr_2RuO_4$ has attracted the most attention as a candidate material but experimental observations from different groups are still contradictory[13-15] and the order parameter of this system has not yet been established. In this context, uranium-based heavy fermion compounds such as $UPt_3$[16], $UGe_2$[17], $URhGe$[18], and $UCoGe$[19] are particularly interesting since the coexistence of magnetism and superconductivity is highly promising for realizing a spin-triplet state[4-10]. However, the extremely low superconducting (SC) transition temperatures $T_{sc}$ and a SC phase that is either buried deep within the ferromagnetic phase or that appears only under high pressure have hampered a deeper investigation of the nature of the SC order parameter[10]. Recently, superconductivity with a transition temperature $T_{sc}$ = 1.6K was discovered in the heavy fermion compound, $UTe_2$[11]. Experiments report several striking properties[11,20-24]: a $^{125}$Te Knight shift in nuclear magnetic resonance measurements which is temperature independent across $T_{sc}$[11]; the upper critical field ($H_{c2}$) largely exceeding the Pauli limit[11, 20]; two reentrant SC phases in very high magnetic fields[21,22]; and strong nearly critical ferromagnetic fluctuations[11,23,24]. These striking observations suggest a spin-triplet pairing scenario. Moreover, the specific heat ($C/T$) drops only to about half of its normal state value in the SC state, independent of sample quality[11,20]. The large residual Sommerfeld γ-coefficient has led to speculations of exotic nonunitary pairing. The discovery of superconductivity in $UTe_2$ presents a novel candidate system to study spin-triplet superconductivity and its interplay with the Kondo effect, and potentially provides a unique opportunity to observe the long-sought Majorana modes predicted to exist on the surfaces of chiral spin-triplet superconductors[2].

In this work we report the investigation of $UTe_2$ by scanning tunneling microscopy/spectroscopy (STM/STS), which reveals direct evidence for chiral spin-triplet pairing. $UTe_2$ crystallizes into an orthorhombic structure with a space group *Immm* (No. 71)[25,26]. Sample quality and bulk superconductivity were confirmed by Laue diffraction, specific heat measurements, and STM on the same sample (Extended Data Fig. 1). After considering the bond distance between U, Te1, and Te2 (Fig.1a and Extended Data Fig. 2), one easy-cleave surface is the (011) plane. As seen from the top view, the (011) plane consists of chains of both Te1 and Te2, as well as U-atoms along the [100] direction. Consequently, both Te2 and Te1 are visible in the STM topography (see Fig. 1b) and form one dimensional (1D) chains along the *a*-axis. In this cleave plane, the Te1 rows are slightly higher than the Te2 rows (Te2 being 6.5 pm below the Te1 layer), which is seen in the topography as alternating rows of bright and dark chains.

Since the normal state of UTe$_2$ is a heavy fermion metal, our first task is to look for signatures of the underlying Kondo effect. Figure 1c shows an averaged d$I$/d$V$ spectrum obtained on a clean surface. In addition to an overall V-shape in the LDOS that extents up to few hundred meV (Extended Data Fig. 3), we observe a low energy feature with a characteristic Fano lineshape, first observed in STM data on a single Kondo impurity[27]. However, as with other heavy fermion materials, UTe$_2$ constitutes a Kondo lattice system, developing a nearly flat band of almost localized heavy quasiparticle states near $E_F$ at low temperatures, with a coherence temperature of roughly 30 K[20]. This nearly localized band is likely responsible for the resonance feature[28] shown in Fig. 1c. A more detailed analysis of the resonance line shape is included later in this paper as well as in the Extended Data Fig. 3. Similar features have been observed in YbRh$_2$Si$_2$[29], CeCoIn$_5$[30], and SmB$_6$[31], which were interpreted in terms of a Kondo lattice peak.

Zooming into a much smaller, 1 meV energy range, we observe the suppression of the LDOS with symmetric coherence peaks located around ±0.25meV, as shown in Fig. 1d. We associate this feature with the SC gap, since it is suppressed both with increasing temperature and magnetic fields, disappearing at the bulk $T_{SC}$ (Fig. 1d and Extended Data Fig. 1) and upon approaching the upper critical field ($H_{c2}$) (Extended Data Fig. 4). However, the tunneling conductance does not go to zero as would be expected for a fully-gapped superconductor, but instead displays a rather large density of gapless excitations. By comparing the spectra to other materials with a similar $T_{sc}$ we can show that the residual LDOS at zero bias is too large to be ascribed to a thermal smearing effect (Extended Data Fig. 4). The large residual density of states is also confirmed by spectra obtained with a SC tip (Extended Data Fig. 4). This residual density of states may arise both due to a background density of unpaired electrons as well as due to low-energy quasiparticle excitations. While a nonunitary pairing state[11] or a hidden order[20] may be invoked to explain a remnant density of unpaired electrons, there is another intriguing possibility that could generate states inside the SC gap. Similar to proposals for candidate spin-triplet superconductors UPt$_3$[32,33] and Sr$_2$RuO$_4$[34-36], chiral dispersing surface states with sub-gap energies are constrained to exist by topology in superconductors with a non-zero Chern number[3]. The experimental signature of such chiral surface states would be a non-vanishing DOS inside the SC gap[36-39]. We will revisit this possibility later.

A distinct property of strongly correlated materials is the fierce competition between various interactions, especially in the quantum critical regime[40]. UTe$_2$ is believed to be located in the vicinity of a quantum critical point as evidenced by the critical scaling of the magnetization[11]. Unlike other uranium-based superconductors, UTe$_2$ exhibits a paramagnetic ground state hosting both strong ferromagnetic fluctuations and Kondo coupling[23,24]. Utilizing STM, we can directly image the competition or coexistence of different order parameters in real space. Figure 2d and 2e show line cuts across the surface from Te1 to Te2 in the energy range of the Kondo resonance and the SC gap, respectively. We find that both the Kondo lattice peak and SC gap show real space modulations within the unit cell. d$I$/d$V$ maps at -6 mV and 0 mV (Fig. 2 b and c) affirm that these periodic modulations are a general feature of this system. Intriguingly, a comparison between

Fig. 2 a, b, and c indicates that while the modulation wave vectors of the LDOS at -6 mV and 0 mV are locked to the atomic structure of Te chains, the height of the peak we attribute to the Kondo resonance and the depth of the SC gap are anticorrelated.

To further characterize the behavior of the Kondo resonance, we fit the peak-dip feature by a Fano lineshape[41]: $dI/dV(V) \propto \frac{((V+E_0)/\Gamma+q_K)^2}{1+((V+E_0)/\Gamma)^2}$. This formula describes tunneling between a localized many-body states at $E_0$ and an itinerant continuum, where $\Gamma$ is proportional to the width of the Kondo resonance and $q_K$ is the Fano parameter. The results of fitting presented in Extended Data Fig. 5 and Table S1 reveal a monotonic change in $\Gamma$ and $|q_K|$ from Te1 to Te2 sites. This seems natural since the nearest U-Te1 distance is shorter than U-Te2 distance, so the coupling to the U $f$-electrons is stronger for the Te1 sites. A similar spatial modulation of the Kondo resonance was also observed in URu$_2$Si$_2$, and thought to be connected with the unsolved mystery of the "hidden order" phase[42,43]. Interestingly, the magnitude of the SC gap, Δ, also shows a modulation from Te1 to Te2, being enhanced by about 2.5 times (Extended Data Fig. 6). A short-range modulation of the SC gap in zero magnetic field is rarely observed, except in the case of a pair density wave seen in cuprate superconductors such as La$_{2-x}$Ba$_x$CuO$_4$[44] and Bi$_2$Sr$_2$CaCuO$_{8+x}$[45]. Our gap modulation does not break any of the lattice symmetries; nevertheless, the strong modulation of Kondo resonance and superconductivity imply a crucial role of atomic orbitals in mediating both effects[46].

Our next step it to elucidate the nature of the SC order parameter in UTe$_2$. To do this we exploit the possibility that step edges could provide crucial information on the pairing symmetry of unconventional superconductors[36-38,47]. Our d$I$/d$V$ measurements at single step edges along the $a$-axis (Fig. 3d and e) reveal an unusual asymmetric lineshape (Fig. 3g and h); a peak-dip feature that conspicuously breaks the particle-hole (p-h) symmetry expected for Bogoliubov quasiparticles. As shown in Fig. 3g and h, asymmetric peaks appear inside the SC gap, either above or below $E_F$, and are accompanied by corresponding dips at the opposite energy. We find that step edges with a normal of the side surface in the [01-1] direction show peaks ~$E_{A-}$ = -0.2 mV (spectra color coded in blue) and those with an opposite normal vector have peaks ~$E_{A+}$ = +0.2 mV (spectra color coded in red) (see Fig. 3 a and b). The peak positions only depend on the surface normal and are insensitive to local features such as the terminating atom or the distance between the steps, (Extended Data Fig. 7). From data on more than 30 step edges on four different samples, we determine that the observed "chiral" phenomenology is universal. The combined data indicate that the asymmetry is not controlled by microscopic details but is rather controlled by a global symmetry of the system.

Additional data show that these "chiral" features are intimately tied to superconductivity. First, the peak/dip energy scale lies within the SC gap (light yellow region in Fig. 4a). Second, our temperature dependent measurements (Fig. 4b) show that the resonances attenuate steadily with increasing temperature, disappearing just around $T_{sc}$. Third, magnetic field dependent d$I$/d$V$ maps along the step edges reveal that the resonances become steadily weaker upon application of a magnetic field and become difficult to detect around $H_{c2}$ (≈10T at 0.3K)[21] (Fig. 4d). These additional data together provide strong evidence that the dI/dV features are a consequence of superconductivity[33] with the asymmetry being controlled by a vector associated with the SC order. The consistent transport evidence for triplet pairing naturally suggests that this vector may be the chiral-axis of a chiral SC state which, given recent thermal conductivity measurements showing evidence for point nodes along $a$-axis[48], would lie along the $a$-axis.

Before proceeding further, it is necessary to address the question of whether a chiral state of the right symmetry is possible for UTe$_2$. Since, the crystal structure is orthorhombic, to obtain an order parameter with two degenerate components one requires either an accidental degeneracy or the realization of a non-unitary (spin polarized) pairing state induced by the vicinity of ferromagnetic order[49]. While pinning down the exact order parameter requires a better knowledge of the band structure, one possible scenario consistent with the available information is as follows. The magnetic properties of UTe$_2$ show a single easy axis parallel to the crystalline $a$-axis, while the other two directions are hard-axes[11,50]. The combination of two non-degenerate pairing states ($d_{1,2}(k)$) can couple to a local spin magnetic moment $M$ along the easy axis, $d_1(k)$= $\Delta_{11} x k_y + \Delta_{12} y k_x$ and $d_2(k)$= $\Delta_{21} x k_z + \Delta_{22} z k_x$ with $M \approx i\langle d_1^* \times d_2\rangle_k$ (here, $d_{1,2}$ are the d-vectors involved in the spin triplet pairing state, $\Delta_{ij}$ is the gap amplitude). This requires a relative phase of π/2 between $d_1$ and $d_2$ which would then lead to a chiral SC phase with the chiral axis parallel to the a-axis. Such a chiral superconductor would be topologically non-trivial with broken time-reversal symmetry.

With the chiral axis along the $a$-axis, the top and/or side surfaces are expected to host topologically protected states with sub-gap energies (also known as surface Andreev bound states) and a linear dispersion along certain directions parallel to the surfaces[39,51,52]. These surface states are particle like for one sign of this momentum $k$ and hole like for the other (Fig.4e). As theoretically discussed by Kobayashi et al.[39], such surface states exist, in principle, as long as the chiral axis of the SC bulk phase is not normal to surface. These chiral states are likely responsible for the in-gap features at the step edges and may be responsible for the significant non-zero low-energy density of states seen in our spectra. Our next step is to understand the lineshape of the tunneling spectra at the step edges.

An explanation of asymmetric lineshapes requires two ingredients: chiral edge states combined with a selective tunneling process. One possible scenario is a momentum selective tunneling picture based on symmetry considerations as shown schematically in Fig.4e and Extended Data Fig. 8. Since the STM tunnel current generally involves an average over all directions of momenta into the surface, tunneling into the chiral surface states on planar surfaces far from the step edges would lead to a $p$-$h$ symmetric tunneling spectrum. However, step edges as realized on our cleaved surfaces locally break spatial reflection symmetry in a way that constrains the tunneling processes. The electrons can now tunnel with a finite mean momentum in the direction parallel to the surface, as shown in Extended Data Fig. 8c. If this mean momentum lies parallel to the momentum of the chiral dispersion of the in-gap surface states, we obtain a selectivity of tunneling into the chiral sub-gap states and we probe the only the electron-like (hole-like) states at negative (positive) energies for the [01-1] ([0-11]) step. In this way the tunneling spectrum would be asymmetric in the voltage, as observed. Interestingly, our data on 45°-step edges (Fig. 3i) show a bound state centered around the Fermi energy. This indicates a reduced effect of momentum selective tunneling since at a 45° angle off from the chiral dispersion direction the tunneling electrons would access positive- and negative-momenta more evenly. Moreover, since bound states at edges are known to depend on the detailed SC gap structure[39], the observed peak at the 45°-step edges (Fig. 3i), provides further constrains for modeling the SC order parameter. The key point here is that asymmetric d$I$/d$V$-curves cannot be explained without invoking both a tunneling related mechanism as well as the presence of chiral edge stages. However, more a detailed theoretical model is necessary to understand whether we need to additionally consider other tunneling processes for chiral states at step edges in a superconductor.

Our data on UTe$_2$ show many intriguing features in this heavy fermion superconductor and suggest a chiral SC state. The anti-correlated intra unit-cell modulation of the Kondo resonance and superconductivity is an indication of competing interactions in the quantum critical region and further theoretical modeling is needed to understand the origins and full implications. Moreover, we provide evidence for the presence of chiral modes existing inside the SC gap, which are expected to occur on the surface of topological/Weyl superconductors. In combination with the temperature independent Knight shift across $T_{SC}$, the very large upper critical field of ~40 T, and evidence for ferromagnetic fluctuations, our data are highly suggestive of UTe$_2$ being a triplet-chiral superconductor. Finally, a 3D chiral superconductor provides a natural platform for the solid-state realization and exploration of Majorana fermions.

**Methods**

Single crystals of UTe$_2$ were synthesized with the chemical vapor transport method using iodine as the transport agent. Elements of U and Te with atomic ratio 2:3 were sealed at one end of an evacuated quartz tube, together with 3 mg/cm3 iodine. The ampoule was gradually heated up and hold in the temperature gradient of 1060/1000 °C for 7 days, with source materials at the hot end. It was then furnace cooled to the room temperature, and mm size crystals were accumulated at the other end. The crystal orientation was determined by Laue diffraction. Samples were all cleaved *in situ* at ~77 K and in ultra-high vacuum chamber. After cleaving, samples are directly transferred to the STM. All STM measurements were performed below an an instrument temperature of 4 K using annealed tungsten tips. *dI/dV* spectra were collected using a standard lock-in technique at a frequency of 913 Hz.

**Figure 1**

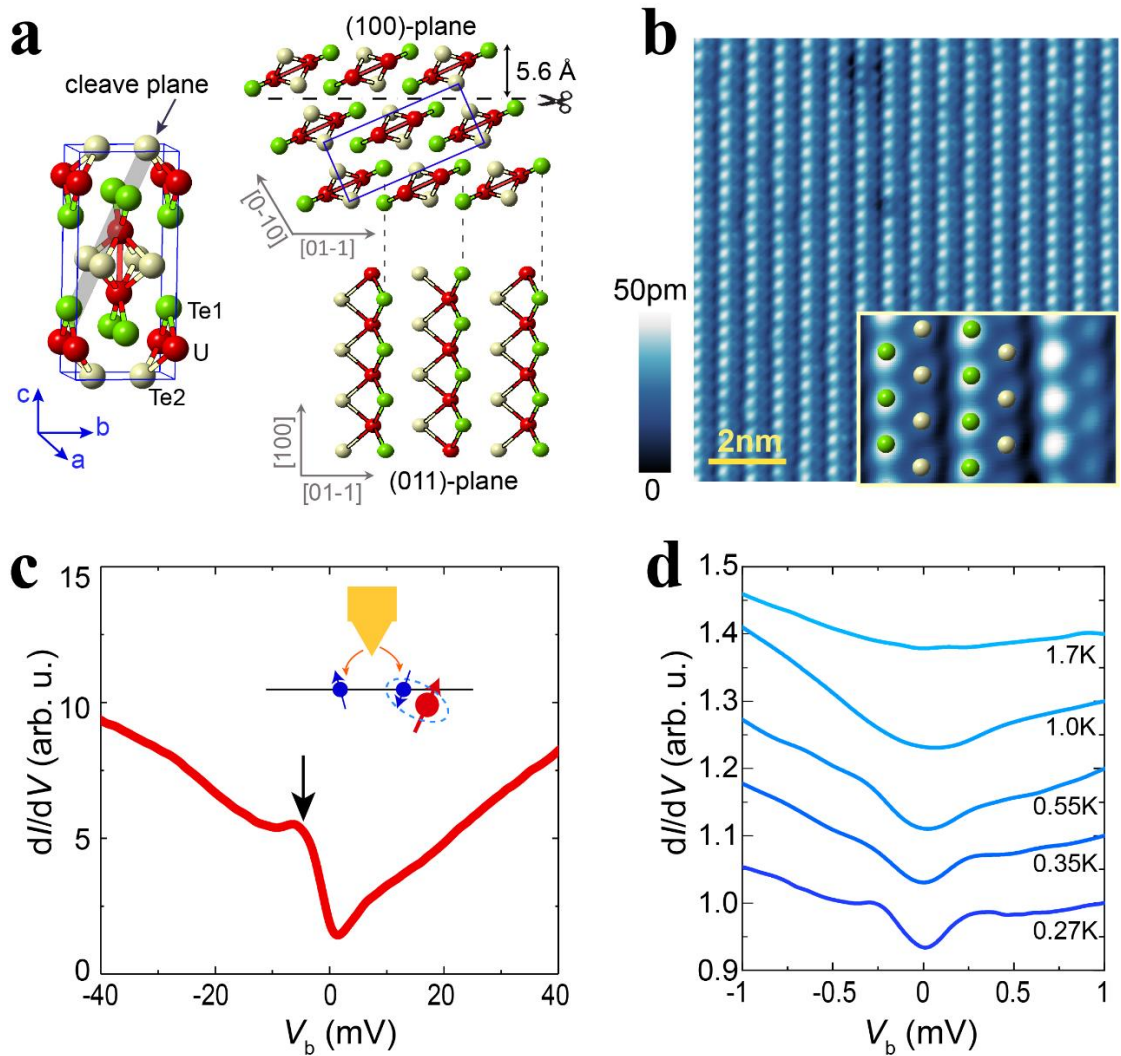

**Figure 1 | Crystal structure and spectroscopic properties of UTe$_2$. a**, Unit cell of UTe$_2$. Our data reveal that the (011) plane is the easy-cleave plane. **b**, STM topography of the exposed (011) plane of UTe$_2$. Te1-chains (bright dots) and Te2-chains (dim dots) can be distinguished. The inset shows the position of Te1 and Te2 atoms. **c**, d$I$/d$V$-curve of UTe$_2$ showing the Kondo resonance ($I_{set}$=150 pA, $V_b$= 50 mV). A typical asymmetric Fano lineshape is observed, the black arrow marks the Kondo lattice peak at -6 mV. The inset sketches the cotunneling effect which gives rise to the Fano lineshape. **d**, Temperature dependent d$I$/d$V$-curves of UTe$_2$ with ± 1 mV, the SC gap with symmetric coherence peaks is visible at low temperature. With increasing temperature, the gap is gradually suppressed and completely flattened at 1.7 K. Spectra are taken at 0 Tesla. ($I_{set}$= 50 pA, $V_b$= 2 mV, $V_{mod}$= 60 µV).

**Figure 2**

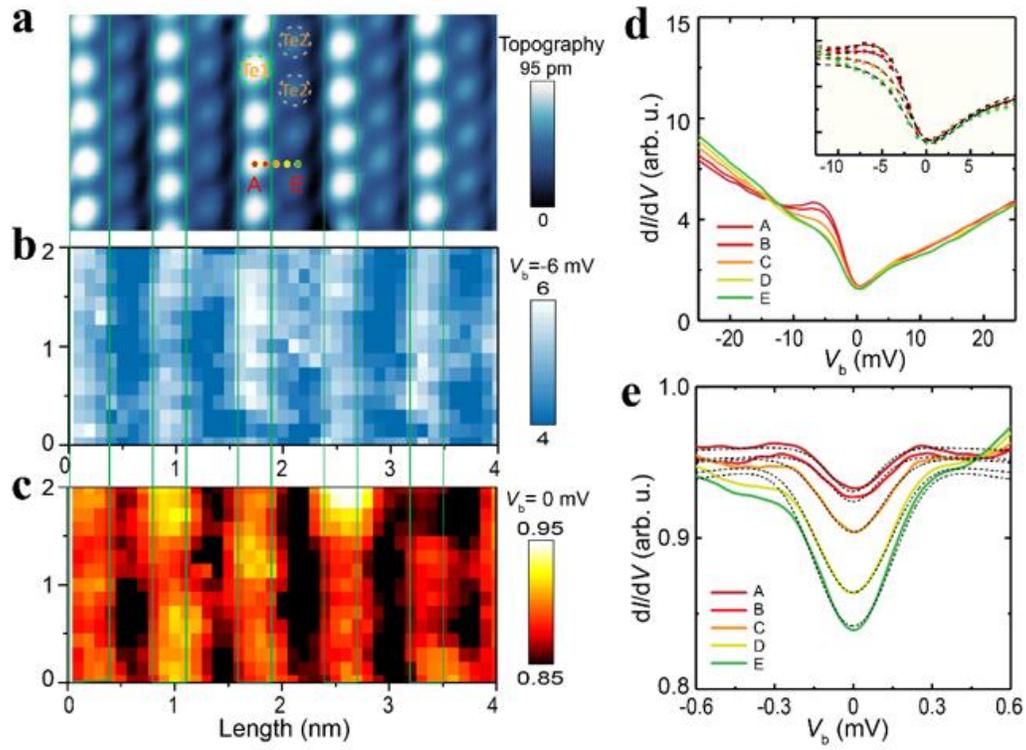

**Figure 2 | Intra-unit cell spatial modulation of Kondo resonance and superconductivity. a**, 2x4 nm$^2$ topography of UTe$_2$. **b**, **c**, d$I$/d$V$-map in the same area as topography at -6 mV (**b**) and 0 mV (**c**). **d,e** d$I$/d$V$-curves measured at the points denoted in **a**. Inset to d shows the experimental data (dots) and the corresponding fits to Fano lineshapes (black dashed line). The black dashed lines in **e** are fittings to Dynes function. Detailed information on the fit parameters is included in the extended data.

**Figure 3**

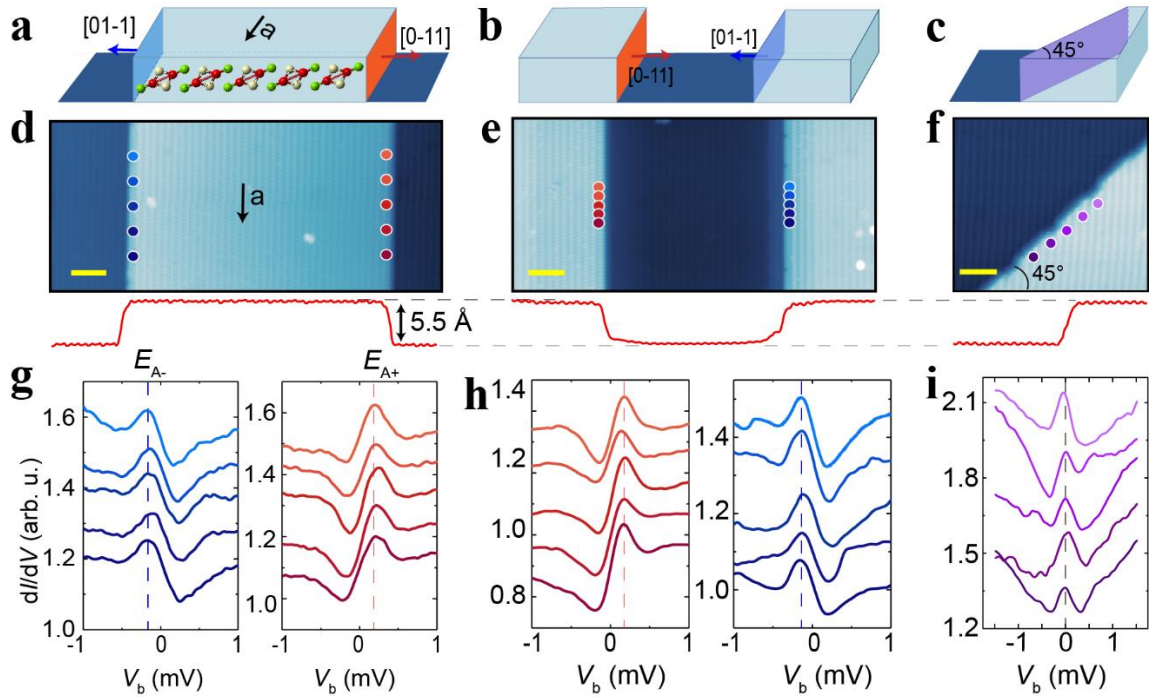

**Figure 3 | "Chiral" in-gap states in UTe$_2$. a-c**, Schematic of surfaces with a terrace, a trench, and a diagonal step-edge (oriented at 45° with respect to the *a*-axis). **d-f**, Topographies of UTe$_2$ with terrace, trench and 45° step-edge. The length of the yellow scale bar is 4 nm. The linecuts of the topographies are shown below and reveal step heights of 5.5 Å corresponding to single-steps. **g-i**, d$I$/d$V$ spectra measured at the positions marked by dots in the topographies. The dashed lines mark the peak positions. Curves are equally shifted for clarity. Data sets were obtained at 0.3 K and zero field.

**Figure 4**

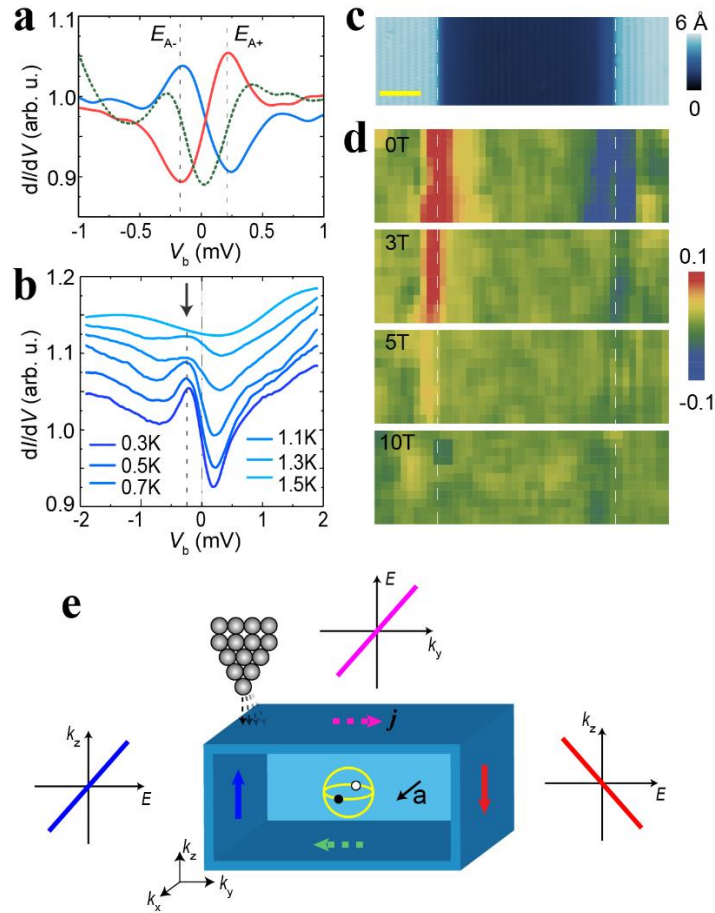

**Figure 4 | Phenomenology of chiral edge states. a**, A comparison of the d$I$/d$V$-curve obtained on a clean surface (green dashed line), [01-1] (blue curve) and [0-11] (red curve) step-edge. **b**, Temperature dependent d$I$/d$V$-curves measured on the *n*-type step-edge. **c**, topography of UTe$_2$ surface (35x10 nm$^2$) with a wide trench. **d**, magnetic field dependent d$I$/d$V$-maps on this area are measured from 0T to 10T. The maps shown depict the density of states differences between positive and negative bias voltages at each location, i.e., d$I$/d$V$(r,$V_b$)-d$I$/d$V$(r,-$V_b$) with $V_b$= 0.2 mV. The successive maps at higher fields show that the "chiral" bound states gradually disappear with increasing magnetic field. **e** A schematic of one possible realization a chiral superconductor in UTe$_2$ with its chiral axis along **a**, showing the resulting chiral surface currents (*j*). *The* colors represent the winding phase of the SC order parameter from 0 to 2π. A simplified spherical Fermi surface with point nodes (black and white dots) along *a*-axis is placed in the center. The linear dispersion of the quasiparticles is plotted close to the corresponding surfaces. The STM tip is shown located on one edge.

**Extended data figures and tables for**

**Microscopic evidence for a chiral superconducting order parameter in the heavy fermion superconductor UTe$_2$**

*Lin Jiao[1], Sean Howard[1], Sheng Ran[2,3], Zhenyu Wang[1], Jorge Olivares Rodriguez[1], Manfred Sigrist[4], Ziqiang Wang[5], Nicholas Butch[2,3], Vidya Madhavan[1]*

[1]Department of Physics and Frederick Seitz Materials Research Laboratory, University of Illinois Urbana-Champaign, Urbana, Illinois 61801, USA
[2]NIST Center for Neutron Research, National Institute of Standards and Technology, Gaithersburg, MD 20899, USA
[3]Center for Nanophysics and Advanced Materials, Department of Physics, University of Maryland, College Park, Maryland 20742, USA
[4]Institute for Theoretical Physics, ETH Zurich, 8093 Zurich, Switzerland
[5]Department of Physics, Boston College, Chestnut Hill, Massachusetts 02467, USA

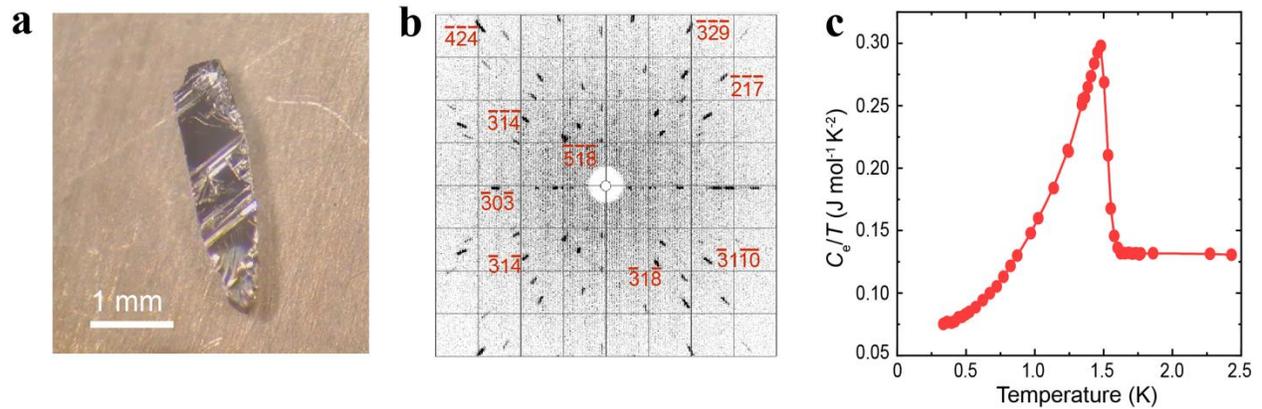

**Extended Data Fig. 1| High quality of UTe$_2$ single crystal. a**, Optic photo of one UTe$_2$ single crystal after being cut along a-aixs. The exposed (011) surface is flat and shiny with several step-edges along the a-axis. **b**, Laue diffraction pattern of one UTe$_2$ single crystal, with selected (hkl) index marked. The same crystal was used for the specific heat and STS measurements presented in c and d. **c**, specific heat of UTe$_2$ shows a pronounce single jump around 1.6 K. The specific heat jump at $T_{sc}$ is particularly large, indicating bulk superconductivity of heavy electrons. The specific heat and temperature dependent STS measurement (shown in Fig. 1d) are conducted on the same sample, indicating well-defined superconducting phase of UTe$_2$.

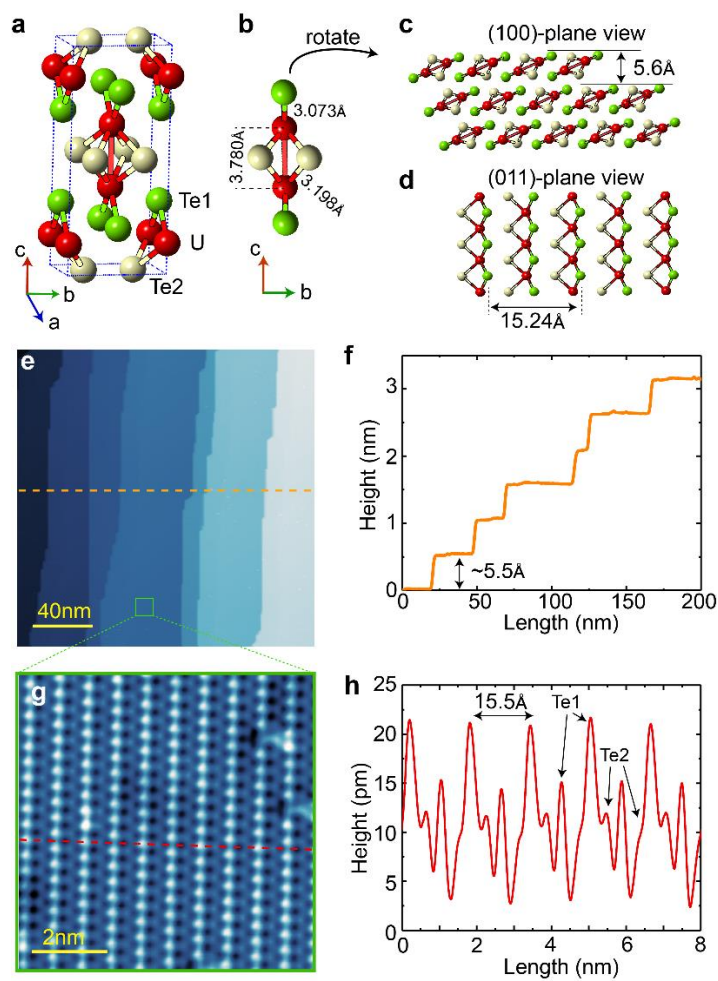

**Extended Data Fig. 2| Crystal structure and cleave plane of UTe$_2$. a**, Crystal structure of UTe$_2$. The lattice parameters are $a$ = 4.161 Å, $b$ = 6.122 Å, and $c$ = 13.955 Å. **b**, the primitive cell of UTe$_2$ with the nearest bond distances between U and Te1/Te2 are denoted in the figure. **c**, (100) view of the crystal showing the cleave plane which is between the primitive cells and breaks the next nearest U-Te1 bond (3.201 Å). **d**, Top view of the cleaved (011) plane. **e**, A typical STM topography of the sample with multiple random distributed step-edges. **f**, Height profile of in the center of **c**. The height of the step-edge is ~5.5 Å. **g**, A 8x8 nm$^2$ topography which shows chains of Te1 and Te2 along $a$-axis. **h**, Height profile in **g** along the red dashed line. The periodicity of the height profile fits very well with lattice parameters in **d**. The step heights and the atomic spacings are consistent with the identification of the cleaved surface with the (011)-plane.

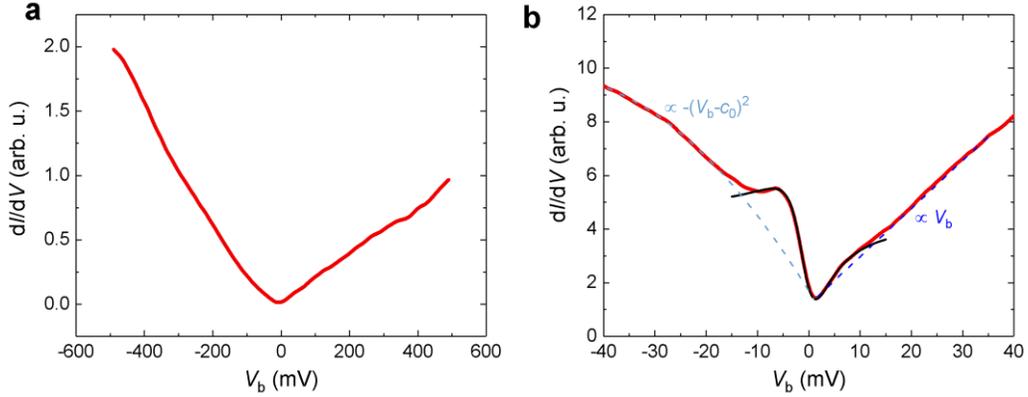

**Extended Data Fig. 3| Large energy LDOS of UTe$_2$. a**. d$I$/d$V$ curve of UTe$_2$ obtained at 0.3K from 400 mV to -400 mV. The curve follows a V-shape at higher energies and drops close to zero around zero bias. The almost zero DOS at $E_F$ suggests that this material is close to being a Kondo insulator. **b**, A fit of the spatially averaged d$I$/d$V$-curve to the Fano model: $[dI/dV(V) \propto \frac{((V+E_0)/\Gamma+q)^2}{1+((V+E_0)/\Gamma)^2}]$. Red and black curves are the raw data and the fitted curve, respectively. The fitting parameters are $q$ = 0.57 (2), $E_0$ = 0.70(8) meV and $\Gamma$ = 3.60 (6) meV. Blue dashed lines are simple (linear for $V_b$>0 and quadratic for $V_b$<0) extrapolation of the V-shaped background, which are subtracted from spectra as shown in Extended Data Fig.3b for further analysis.

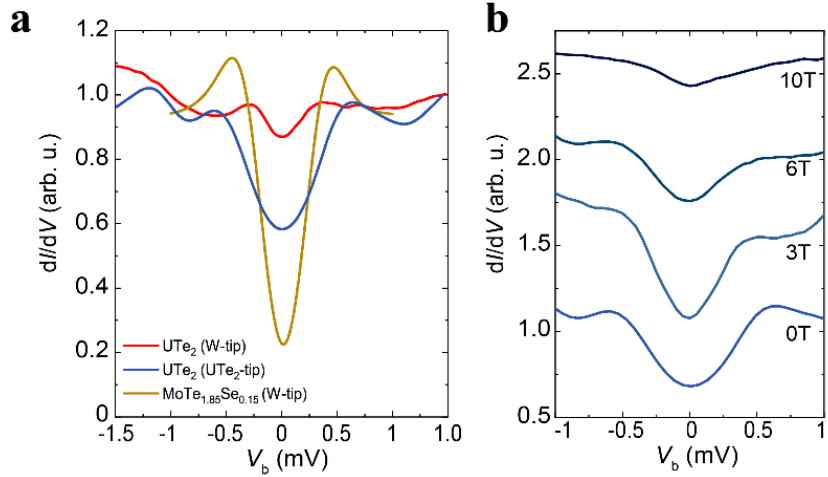

**Extended Data Fig. 4| Phenomenology of superconducting gap of UTe$_2$. a**, Comparison of the superconducting gap structure of UTe$_2$ obtained with a tungsten (W)-tip and UTe$_2$ tip as well as data on superconducting MoTe$_{1.85}$Se$_{0.15}$ for comparison. The spectra were all obtained using the same STM system at ~0.3 K without an external magnetic field. MoTe$_{1.85}$Se$_{0.15}$ possesses a $T_{sc}$ ≈ 2.2K and gap Δ ≈ 0.3 meV, slightly larger than UTe$_2$. In UTe$_2$, the LDOS at zero bias only drops 5 to 10% of its normal state DOS, while it drops by ~80% in MoTe$_{1.85}$Se$_{0.15}$. Even in the data measured by UTe$_2$-tip which shows a larger gap than MoTe$_{1.85}$Se$_{0.15}$, the LDOS drop is smaller. Since all the data were collected with similar tunneling parameters by the same STM system, the large residual LDOS in UTe$_2$ is unlikely to be a simple thermal broadening effect or due to some external artifact. **b**, Magnetic field dependent d$I$/d$V$-curves at 0.3K. As shown in the main text, the intrinsic superconducting gap is small. Therefore, tracking the superconducting gap structure in high magnetic field is challenging. Instead, we present the data obtained with a superconducting tip (a W-tip with UTe$_2$ flake at the apex). The figure shows that the superconducting gap is continuously suppressed up to 10 T. Since $T_{sc}$ is around 1.6 K, this result suggests a large upper critical field of UTe$_2$, exceeding the Pauli limit of 1.86$T_c$. Meanwhile, the derived upper critical field of ~10 T is consistent with transport measurements with the field perpendicular to the [011] plane.

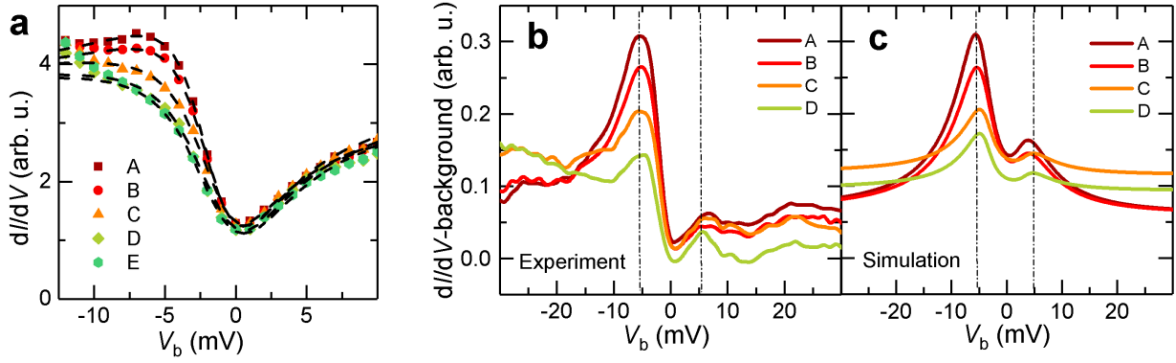

**Extended Data Fig. 5 | Analyses of the Kondo lattice peak. a**, Fits to the Kondo resonance feature obtained as a function of position from Te1 (site A) to Te2 (site E) (see main text). Curves are fitted to Fano line shape. The fitting parameters are summarized in Extended Data Table1. The Kondo temperature $T_K$ can be derived from $\Gamma = 2k_B T_K$. As shown in the table, $T_K$ varies from ~19.6K to 26K. **b,** the LDOS (d$I$/d$V$) obtained by subtracting a V-shaped background (dashed blue lines in Extended Data Fig. 2). **c**, simulation of the d$I$/d$V$ by a Kondo lattice model with quantum cotunneling effect. The model used here is proposed by Maltseva et al., [1]:

$$\frac{dI}{dV}(V) = \mathrm{IM}\left[\left(1 + \frac{v*q}{V - i*\gamma - E_0}\right)^2 * \mathrm{Log}\left[\frac{V - i*\gamma + D1 - \frac{v^2}{V - i*\gamma - E_0}}{V - i*\gamma - D2 - \frac{v^2}{V - i*\gamma - E_0}}\right] + \frac{(D1 + D2)*q^2}{V - i*\gamma - E_0}\right]$$

where the definitions of $E_0$, $q$ are the same as those in the Fano model, −D1 and D2 are the energy level of the lower and upper conduction band edges, $v$ is the hybridization amplitude, γ is the self-energy. For simplicity, we take the same $E_0$, $q$ from table 2. D1 ≈ 3 eV is obtained from angle-resolved photoemission spectroscopy. Then the remaining free parameters are D2, $v$ and γ. Table 2 presents the fitting parameters. Given the Kondo hybridization gap $\Delta_K = 2v^2/(D1+D2)$, we also calculated the hybridization gap size in Extended Data Table 2, which is comparable to $\Gamma$ in Extended Data Table 1.

**Extended Data Table 1|** Fitting results to Fano line shape for the Kondo features.

| site | q | $E_0$ (meV) | $\Gamma$ (meV) | $T_K$ (K) |
|---|---|---|---|---|
| A | **-0.66(2)** | 1.78(7) | 3.37(5) | 19.6(3) |
| B | **-0.59(2)** | 1.54(8) | 3.51(6) | 20.4(3) |
| C | **-0.41(3)** | 1.0(2) | 4.0(1) | 23.2(6) |
| D | **-0.36(4)** | 1.1(3) | 4.4(2) | 26(1) |

**Extended Data Table 2|** Parameters used to simulate d$I$/d$V$ using Kondo cotunneling model.

| site | q | $E_0$ (meV) | -D1 (eV) | D2 (eV) | γ (meV) | v (meV) | $\Delta_K$ (meV) |
|---|---|---|---|---|---|---|---|
| A | **-0.66** | 1.78 | 3 | 5.4 | 2.6 | 115 | 3.15 |
| B | **-0.59** | 1.54 | 3 | 5.4 | 2.8 | 114 | 3.09 |
| C | **-0.41** | 1.0 | 3 | 5.4 | 2.05 | 118 | 3.32 |
| D | **-0.36** | 1.1 | 3 | 5.4 | 1.85 | 118 | 3.32 |

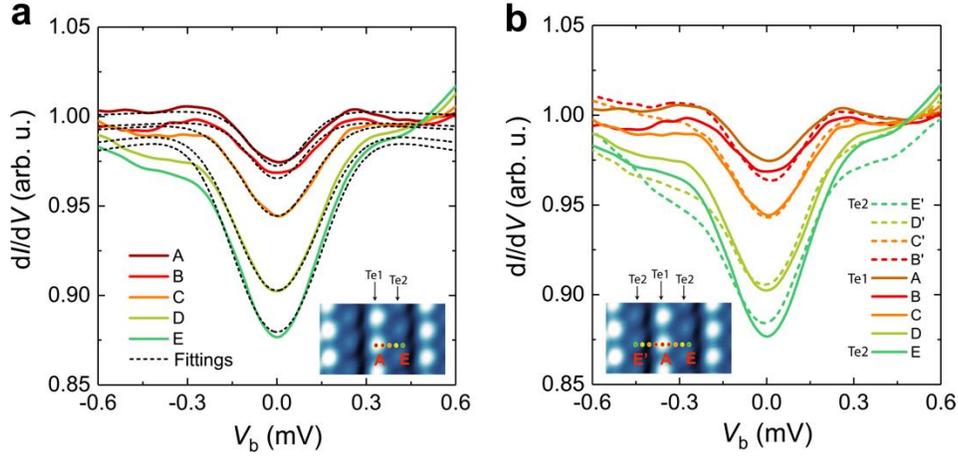

**Extended Data Fig. 6| Analyses of the superconducting gap. a**, Fits to a series of superconducting gaps obtained at positions shown in the inset as we move from Te1 (site A) to Te2 (site E). The d$I$/d$V$-curves are fitted to the Dynes function with the thermal broadening effect included [2,3]: $\frac{dI}{dV}(V) = N_0 \int \mathrm{Re}\left[\frac{V+\omega+iG}{\sqrt{(V+\omega+iG)^2-\Delta(\theta,\phi)^2}}\right]\left(-\frac{\partial f}{\partial \omega}\right)d\omega + N_u$, where $f$ is the Fermi-Dirac distribution function at 0.3 K, $N_0$ is proportional to the LDOS in the normal states, $N_u$ is related to the residual LDOS dominated by unpaired electrons which is set at 0.5 based on the specific heat data, $G$ quantifies the effect of the pair-breaking processes, which is related to the quasiparticle lifetime. The most important parameter here is the superconducting gap function $\Delta(\theta)$. Here, we tried both s-wave gap $\Delta(\theta,\phi)= \Delta_0$ and the proposed spin triplet $p_x+ip_y$ gap $\Delta(\theta,\phi) = \Delta_0|\hat{k}_x + i\hat{k}_y|$. As $N_u$ is approximately 50% of $N_0$, the derived gap sizes $\Delta_0$ at each site are similar between s- and p-wave gap function. Fit parameters are summarized in Extended Data Table 3. For both s- and p-wave gap functions, $\Delta_0$ increases about 2.7 times from site A to site E, while $G$ shows much smaller changes. **b**, Spatial variation of the superconducting gap from one Te2 chain to the next Te2 chain. Oscillations of the coherence peak and zero bias LDOS are manifested in the figure.

**Extended Data Table 3| Summary of the analysis of d$I$/d$V$-curves to the s- and p-wave gap functions.**

| | p-wave | | s-wave | |
|---|---|---|---|---|
| site | $\Delta_0$ (meV) | $G$ (meV) | $\Delta_0$ (meV) | $G$ (meV) |
| A | 0.086(4) | 0.17(1) | 0.061(3) | 0.18(1) |
| B | 0.088(5) | 0.18(1) | 0.063(4) | 0.18(1) |
| C | 0.137(5) | 0.21(1) | 0.099(4) | 0.22(1) |
| D | 0.201(8) | 0.22(1) | 0.148(6) | 0.24(1) |
| E | 0.23(1) | 0.21(1) | 0.168(7) | 0.23(1) |

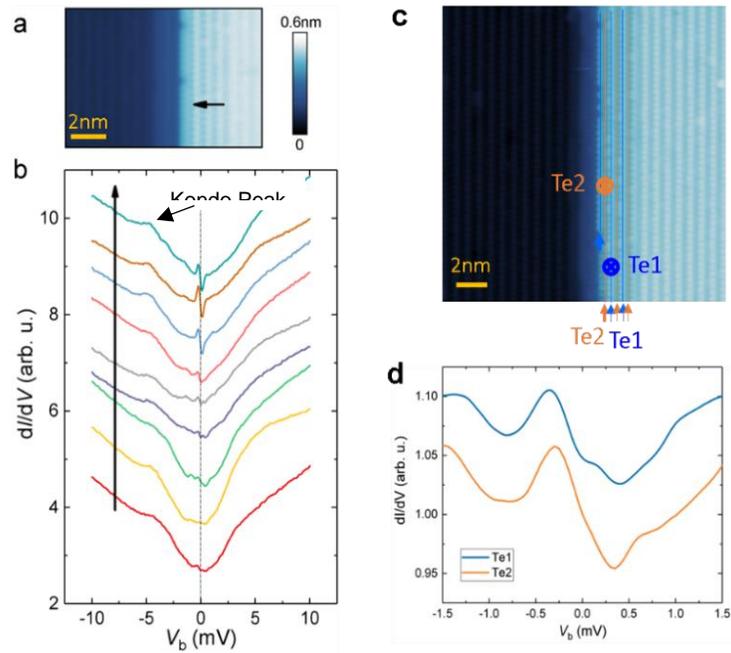

**Extended Data Fig. 7 | Robustness of the Kondo effect and "chirality" of the tunneling current around the step edge**. **a**, Topography with a *n*-type step edge. **b**, d*I*/d*V* curves along the black arrow in **a**, obtained at 9 points equally distributed within a 2 nm length. The results shown in **b** manifest a weak modulation of the Kondo effect, consistent with Extended Data Fig. 5. Importantly, no clear change of the energy level of the Kondo lattice peak is observed around the step-edge when the asymmetric edge states appear. Therefore, the in-gap states coexist with the Kondo effect in UTe$_2$. **c**, Topography showing that a step edge can terminate at either Te1- or Te2-chains. **d**, d*I*/d*V* spectra obtained on Te2 and Te1 sites. The two spectra obtained at Te2- and the Te1-terminated step-edges shown in **c** are similar with the peak always appearing at negative biases, indicating that the "chirality" of the tunneling current is robust against local step-edge termination.

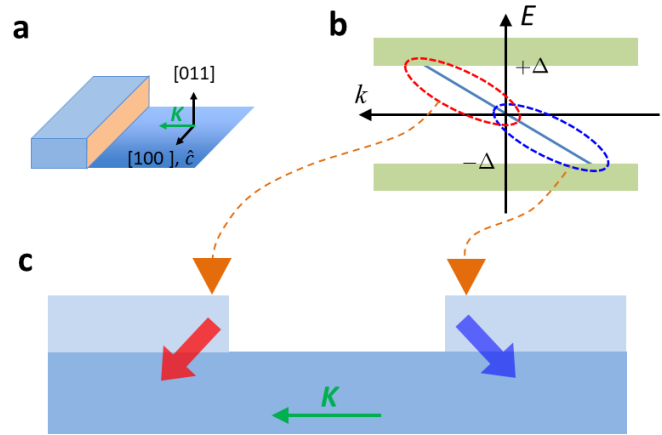

**Extended Data Fig. 8| Schematic of momentum-selective tunneling on the step-edge. a**, Schematic illustration of the sample with orientations of the *a*-axis (which equals to chiral axis, $\hat{c}$), the normal of the cleave surface ([011]) and the momentum ***K*** (spontaneous current direction). **b**, For a chiral superconductor, a quasi-linear dispersive chiral in-gap state (dark blue line) is expected to reside on the surface layers. **c**, On the step edge, the incident momenta distributions of incident tunneling electrons are skewed. In this case, momentum selective tunneling effect are realized close to the step edge, which enable us to tunnel into the states with either +*k* (blue dotted oval) or -*k* (red dotted oval), respectively.